\documentclass[12pt]{article}
\usepackage{graphicx, amssymb, amsbsy, amsmath,color,url,mathtools}
\usepackage{algorithm,algorithmic,natbib,epsfig}
\usepackage{multirow}
\usepackage[export]{adjustbox}
\usepackage{epstopdf}
\usepackage{booktabs}
\usepackage{colortbl,xcolor}
\usepackage{soul}
\usepackage{float}
\usepackage{wrapfig} 
\usepackage[margin=1in]{geometry}
\usepackage{setspace}

\DeclareGraphicsExtensions{.pdf,.eps.gz,.eps,.epsi.gz,.epsi,.ps,.ps.gz}
 
\DeclareGraphicsExtensions{.pdf,.eps.gz,.eps,.epsi.gz,.epsi,.ps,.ps.gz}
\long\def\symbolfootnote[#1]#2{\begingroup%
\def\thefootnote{\fnsymbol{footnote}}\footnote[#1]{#2}\endgroup}

\newtheorem{theorem}{Theorem}%
\newtheorem{lemma}{Lemma}%
\newtheorem{assumption}{{\bf Assumption}}

\newcommand{\dd}{\mathrm{d}}
\newcommand{\bm}{\boldsymbol}
\newcommand{\bbeta}{{\bm\beta}}

\newcommand{\btheta}{{\bm\theta}}

\def\A{{\bm A}}

\def\D{{\bm D}}

\def\c{{\bm c}}

\def\bv{{\bm v}}

\def\1{{\bm 1}}
\def\0{{\bm 0}}
\def\log{\hbox{log}}
\def\squarebox#1{\hbox to #1{\hfill\vbox to #1{\vfill}}}

\def\logit{{\mbox{logit}}}
\def\boxit#1{\vbox{\hrule\hbox{\vrule\kern6pt\vbox{\kern6pt#1\kern6pt}\kern6pt\vrule}\hrule}}

\begin{document}
\title{Mixed-Effect Time-Varying Network Model and Application in Brain Connectivity Analysis}
\vspace{0.5in}
\author{%
{Jingfei Zhang{\small $~^{1}$}, Will Wei Sun {\small $~^{2}$} and Lexin Li {\small $~^{3}$}}%
\vspace{1.6mm}\\
\fontsize{10}{10}\selectfont\itshape
$^{1,2}$\,Department of Management Science, School of Business Administration, \\ 
\fontsize{10}{10}\selectfont\itshape
University of Miami, Miami, FL, 33146. \\
\fontsize{10}{10}\selectfont\itshape
$^{3}$\,Division of Biostatistics, School of Public Health, \\ 
\fontsize{10}{10}\selectfont\itshape
University of California at Berkeley, Berkeley, CA, 94720. \\
}
\date{}
\maketitle

\begin{abstract}
Time-varying networks are fast emerging in a wide range of scientific and business disciplines. Most existing dynamic network models are limited to a single-subject and discrete-time setting. In this article, we propose a mixed-effect multi-subject continuous-time stochastic blockmodel that characterizes the time-varying behavior of the network at the population level, meanwhile taking into account individual subject variability. We develop a multi-step optimization procedure for a constrained stochastic blockmodel estimation, and derive the asymptotic property of the estimator.  We demonstrate the effectiveness of our method through both simulations and an application to a study of brain development in youth.
\end{abstract}


\section{Introduction}
\label{sec:introduction}

The study of networks has recently attracted enormous attention, as they provide a natural characterization of many complex {social, physical and biological systems}. A variety of statistical network models have been developed to analyze data in the form of networks \citep[among many others]{WeiLi2007, bickel2009nonparametric,lib2014network,ZhaoCaiLi2014, Cai2017,zhao2017link}. See also \citet{Kolaczyk2009Book} for a review. To date, much research, however, has focused on static networks, where a single snapshot of the network is observed and modeled. Dynamic networks, where the data consist of a sequence of snapshots of the network that evolves over time, are fast emerging. Examples include brain connectivity networks, gene regulatory networks, protein signaling networks, and terrorist networks. Modeling of dynamic networks has appeared only recently \citep{xu2014dynamic, matias2017statistical, pensky2016dynamic, ZhangCao2017}. Most existing dynamic network models, however, considered a discrete-time and single-subject setting, in which the network observed at multiple time points is based upon the same study subject; the snapshots of the dynamic network are observed on a finite and typically small number of discrete time points. Methodology for modeling dynamic networks in a multi-subject and continuous-time setting remains largely missing. In this article, we develop a new network model to address this question.  

Our motivation is a study of brain development in youth based on functional magnetic resonance imaging (fMRI). In recent years, there has been substantially increasing attention in understanding how brain functional connectivity develops during youth \citep{fair2009functional, supekar2009development}. Brain connectivity reveals synchronization of brain systems via correlations in neurophysiological measures of brain activity, and when measured during resting state, it maps the intrinsic functional architecture of the brain \citep{Varoquaux2013}. Brain connectivity is commonly encoded by a network, with nodes representing brain regions and links representing interaction and coordination between those regions \citep{ChenKang2015, Kang2016, lib2017network}. As such, network based analysis has become a crucial tool in understanding brain functional connectivity. Our data consist of 491 healthy young subjects with age between 7 and 20 years, where age was continuously measured. Each subject received a resting-state fMRI scan, and the image was preprocessed and summarized in the form of a network. The nodes of the network correspond to 264 seed regions of interest in brain, following the cortical parcellation system defined by \citet{Power2011}. The links of the network record partial correlations between pairs of those 264 nodes. Partial correlation has been frequently employed to characterize functional connectivity among brain regions \citep{WangKang2016, XiaLi2017}. Moreover, those 264 nodes have been partitioned into 10 functional modules that correspond to the major resting-state networks defined by \citet{Smith2009}. Each module possesses a relatively autonomous functionality, and complex brain tasks are carried out through coordinated collaborations among those modules. The goal of this study is to investigate how functional connectivity at the whole-brain level varies with age, and in turn to understand how brain develops from childhood to adolescence and then to early adulthood. 

To achieve this goal, we propose a new mixed-effect time-varying stochastic blockmodel. Stochastic blockmodels have been intensively studied in the networks literature \citep{nowicki2001estimation, bickel2009nonparametric, rohe2011spectral, zhao2012consistency}, and there have been some recent studies of dynamic stochastic blockmodels \citep{xu2014dynamic, matias2017statistical,ZhangCao2017}. However, the existing models assume a single-subject and discrete-time setting, and thus are not directly applicable to our data problem. In contrast, by introducing a mixed-effect term in the stochastic blockmodel, our new proposal adapts to the multi-subject setting. It both  characterizes the time-varying behavior of the network at the population level, and accounts for subject-specific variability. Moreover, by modeling the connecting probabilities as functions of the time variable, our method works for the continuous-time setting. In addition, we introduce a shape constraint and a fusion constraint to further regularize and improve our model estimation and interpretation. Both our model choice and the regularization constraints are guided and supported by current neurological findings of brain development studies, as well as evidences from the analysis of the motivating brain development data. The resulting model offers a good balance between model complexity and model flexibility, and is highly interpretable. 

The contributions of our work are twofold. Scientifically, we provide a systematic and rigorous approach to model time-varying networks, and address a class of problems of immense scientific importance. We also remark that, although motivated by a concrete brain development study, our proposed method is not limited to this application alone, but is applicable to a broad range of dynamic network modeling problems. Methodologically, the proposed model is able to both characterize the time-varying nature of the network and account for individual variability. An important contribution of the proposed model is that it can handle both the multi-subject and continuous-time settings. It thus fills an existing gap in the literature on network models. 
For the proposed model, we develop a multi-step optimization procedure to address the challenges in parameter estimation that are introduced by the shape and fusion constraints. We also derive the error bound of the resulting estimator. Our proposal makes a useful addition to the toolbox of network modeling.

The rest of the article is organized as follows. Section \ref{sec:model} develops the mixed-effect time-varying stochastic blockmodel, and introduces the shape and fusion structures on the model parameters. Section \ref{sec:estimation} describes a multi-step procedure for model estimation, and Section \ref{sec:theory} investigates the asymptotic property of the resulting estimator. Section \ref{sec:simulation} demonstrates the effectiveness of the proposed method through simulations, and Section \ref{sec:realdata} analyzes the motivating brain development data. The Supplementary Materials collect all technical proofs.

\section{Model}
\label{sec:model}

Suppose there are $N$ subjects, ordered by a \emph{continuous} time variable. Without loss of generality, we assume the time variable is sorted and normalized in that $0 = r_1 \le r_2 \le \ldots r_{N-1} \le r_N=1$. In our example, the time variable is age, while it can represent other variable that orders the subjects as well. For each subject $i = 1, \ldots, N$, a network $\mathcal{G}_i(\mathcal{V}_i, \mathcal{E}_i)$ is observed, where $\mathcal{V}_i$ denotes the set of nodes and $\mathcal{E}_i$ the set of edges. Here we consider simple networks that are undirected and have no self-loops or multiple edges. Most real world networks, including brain connectivity networks, belong to this class. We also assume that all subjects share the same set of nodes, i.e., $\mathcal{V}_1 = \ldots = \mathcal{V}_N = \mathcal{V}$, and the size of $\mathcal{V}$ is $n$. This assumption is reasonable in our context, as brain images of different subjects are generally mapped to a common reference space. Each network $\mathcal{G}_i$ is uniquely represented by its $n \times n$ adjacency matrix $\bm A_i$, where $A_{ijj'}=1$ if there is an edge between nodes $j$ and $j'$ for subject $i$, and $A_{ijj'}=0$ otherwise, $1 \le j \le j' \le n$. 


We assume the nodes in $\mathcal{V}$ belong to $K$ communities, and denote the community membership of node $j$ as $c_j \in \{1,2,\ldots,K\}$, $j=1,\ldots,n$. We model the entries of the adjacency matrix $\bm{A}_i$ of subject $i$ as independent Bernoulli random variables such that 
\begin{equation} \label{eqn:model}
\begin{split}
A_{ijj'} | r_i=t, c_{j}=k, c_{j'} = l  \sim  \mathrm{Bernoulli}(P_{ikl}), \\ 
g(P_{ikl}) = \theta_{kl}(t) + w_i, \quad w_i \sim \mathrm{Normal}\left( 0, \sigma(t)^2 \right), 
\end{split}
\end{equation}
for $i = 1,\ldots,N, 1 \le j, j' \le n, 1 \le k, l \le K$. In this model, $\theta_{kl}(t)$ is the continuous-time connectivity function that characterizes the population level  between-community connectivity when $k \neq l$, and the within-community connectivity when $k = l$. It is modeled as a function of time $t$, and is treated as a fixed-effect term, as it is commonly shared by all subjects. In our investigation, $\theta_{kl}(t)$ is the primary object of interest. To account for subject-specific variability, we introduce a random-effect term $w_i$ for subject $i$, which is assumed to follow a normal distribution with zero mean. Its variance $\sigma(t)^2$ is also time-varying, which is supported by the existing literature that the subject level variability varies with age \citep{garrett2017age, nomi2017moment}. On the other hand, different communities are assumed to share a common variance function $\sigma^2(t)$. This is a reasonable tradeoff between model complexity and flexibility; it allows one to pool information from different communities to estimate the variance function. 
If there are additional subject-specific covariates, it is straightforward to incorporate those covariates into our model. That is, letting $\bm{x}_i$ denote the vector of covariates from subject $i$, we simply modify \eqref{eqn:model} with $g(P_{ikl}) = \theta_{kl}(t) + \bm{x}_i^\top\bbeta + w_{i}$, where $\bbeta$ is the vector of covariate coefficients. For simplicity, we skip the term $\bm{x}_i^\top\bbeta$ in this article. 

We note that our proposed model is different from those in the literature in several ways \citep{xu2014dynamic,pensky2016dynamic, matias2017statistical,ZhangCao2017}. First, it accommodates multi-subject network data, and includes a random effect term to characterize subject-level variability. 
Second, the proposed model handles the continuous-time setting, while existing methods are limited to the discrete-time setting.
As such, existing time-varying network models cannot be applied when the time variable (e.g. age, observation time,\dots) is continuous.
Furthermore, to increase the model flexibility, we allow both the fixed effect term and the variance of the random effect component to be time-varying.
We call model \eqref{eqn:model} a \emph{mixed-effect time-varying stochastic blockmodel}.

In model \eqref{eqn:model}, $g(\cdot)$ is a known, monotonic, and invertible link function. We set $g(\cdot)$ as the logit link, while other link functions such as the probit link can also be used. 
We remark that, in our model formulation, we assume the network community information is \emph{known}, and it does \emph{not} change over time. This assumption is both plausible and useful scientifically. In brain connectivity analysis, brain regions (nodes) are frequently partitioned into a number of known functional modules (communities). These communities are functionally autonomous, and they jointly influence brain activities. Moreover, although the connectivities within and between the functional modules vary with age, the nodes constituting those modules remain stable over age \citep{yeo2011organization, betzel2014changes}. Focusing on those brain functional modules also greatly facilitates the interpretation, particularly when the brain network changes over time. 

Next, we propose to model the time-varying components $\theta_{kl}(t)$ and $\sigma(t)$ as \emph{piecewise constant} functions. This assumption is desirable for several seasons. Scientifically, numerous studies have found that the development of brain undergoes a mix of periods of rapid growth and plateaus \citep{zielinski2010network, geng2017structural}. Such a pattern can be adequately captured when $\theta_{kl}(t)$ is piecewise constant. Moreover, functions with a simple structure such as piecewise constant enables an easier and clearer interpretation than those involving curvatures. Methodologically, the piecewise constant structure allows one to pool together subjects of similar ages to better estimate the variance of the subject-specific random-effect term $w_i$, which would in turn lead to an improved estimation of the connectivity probabilities. This characteristic is particularly useful, since the individual variability is typically large in brain development studies. Theoretically, we show in Section \ref{sec:theory} that, even when the true connectivity function $\theta_{kl}(t)$ does not admit a piecewise constant form, estimating it with a piecewise constant structure with a diverging number of constant intervals would still achieve a reasonable estimation accuracy. 

\begin{figure}[t!]
\centering
\includegraphics[scale=0.825,trim=4mm 0 0 0]{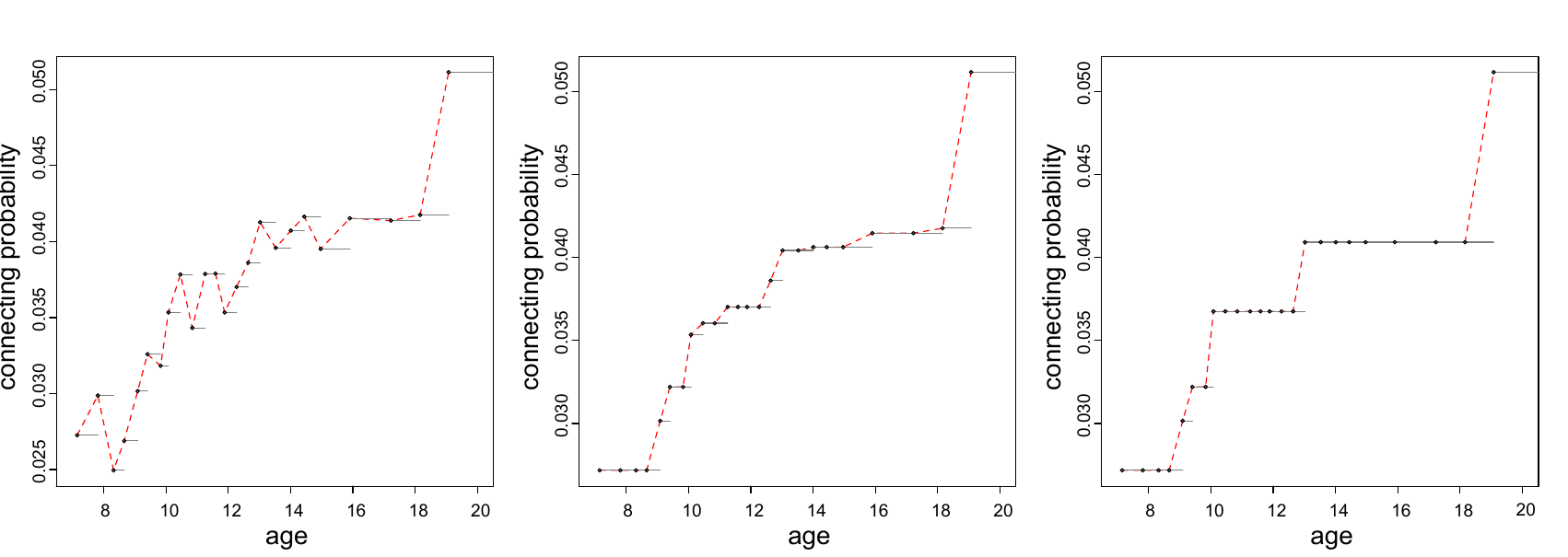}
\caption{Estimated connecting probability between two functional modules, with the endpoints connected with red dashed lines. Left panel: the estimate obtained by minimizing the loss function in \eqref{eqn:loss} with no constraint; middle panel: the estimate with the shape constraint; right panel: the estimate with both the shape constraint and the fusion constraint.} 
\label{fig:c89}
\end{figure}

To model the piecewise constant functions $\theta_{kl}(t)$ and $\sigma(t)$, we partition $(0,1]$ into $S$ intervals; i.e., $(0,1]=\cup_{s=1}^S(t_{s-1},t_s]$ for $0=t_0<t_1<\ldots<t_S=1$. One can either use equal-length intervals, or set each interval to contain the same number of subjects. In Sections \ref{sec:simulation} and \ref{sec:realdata}, we show that our method is not overly sensitive to the number of intervals $S$, in that a similar estimate is obtained as long as $S$ is within a reasonable range. Theoretically, we allow $S$ to go to infinity along with the sample size $N$. This allows our piecewise constant approximation to better capture the time-varying pattern, even if it changes frequently. Following the above partition, we assume that the connectivity function between the communities $k$ and $l$ is constant in $(t_{s-1},t_s]$ and denote it as $\theta_{kl,s}$, and the random-effect variance function is constant in $(t_{s-1},t_s]$ and denote it as $\sigma^2_s$, $s = 1, \ldots, S$. Next we define the vector $\btheta_{kl}=(\theta_{kl,1},\ldots,\theta_{kl,S})^\top \in \mathbb R^{S}$, $\bm{\sigma}=(\sigma_1,\ldots,\sigma_S)^\top \in \mathbb R^{S}$, and the matrix $\bm{\Theta}=( \btheta_{11},\ldots,\btheta_{1K},\btheta_{22},\ldots,\btheta_{2K},\ldots,\btheta_{KK} ) \in \mathcal R^{S \times K(K+1)/2}$. Furthermore, let $\tau_i = \sum_{s=1}^S I(r_i \le t_s)$ denote the interval that contains subject $i$, where $I(\cdot)$ is the indicator function. Write $\bm{A}=\{\bm{A}_1, \ldots, \bm{A}_N\}$, and $\bm{c}=\{c_1,\ldots,c_n\}$, where $\bm{A}_i$ is the adjacency matrix of subject $i$, $i=1,\ldots, N$, and $c_j$ is the community membership of node $j$, $j = 1, \ldots, n$. We then aim to estimate the parameters of interest $\bm{\Theta}$ and $\bm{\sigma}$ by minimizing the following loss function, 
\begin{eqnarray} \label{eqn:loss}
\begin{split}
&&\mathcal L(\bm{\Theta},\bm{\sigma}|\c,\bm{A})=-\log\left\{\prod_{i=1}^{N} \prod_{j=1}^n \prod_{j'>j}^n P_{ic_jc_{j'}}^{A_{ijj'}}\left(1-P_{ic_jc_{j'}}\right)^{(1-A_{ijj'})}\right\}\\
&&= -\sum_{i=1}^{N}\log\left\{\int_{-\infty}^{+\infty}\prod_{k=1}^K\prod_{l\ge k}^K\frac{\left(e^{\theta_{kl,\tau_i} + \, w_{i}}\right)^{n_{ikl}}}{\left(1+e^{\theta_{kl,\tau_i} + \, w_{i}}\right)^{n_{kl}}} \, \phi(w_{i}, \sigma_{\tau_i}^2) \, \mathrm{d} w_{i}\right\},
\end{split}
\end{eqnarray}
where $\phi(\cdot, \cdot)$ is the zero mean normal density function, $n_{ikl}=\sum_{1\le j<j'\le m}A_{ijj'}I(c_j=k,c_{j'}=l)$, $n_k=\sum_{j=1}^n I(c_j=k)$, and $n_{kl}=n_k n_l$, $1\le k\le l\le K$, $i=1,\ldots,n$. Figure \ref{fig:c89}, the left panel, shows the estimated connecting probability $g^{-1}(\theta_{89})$ between the 8th and 9th communities, {\emph{executive control}} and \emph{frontoparietal right}, in our brain development example, after minimizing the loss function \eqref{eqn:loss}. Details on minimization of \eqref{eqn:loss} are given in Section \ref{sec:estimation}. 

It is seen from the plot that the estimated connecting probability roughly follows an increasing trend along with age; however, it is not strictly monotonic, with small fluctuations from approximately 9 to 16 years old. Next we introduce a shape constraint to the loss function \eqref{eqn:loss}, in that $\theta_{kl}(t)$ follows a certain shaped trajectory, such as monotonically increasing, monotonically decreasing, unimodal, and inverse unimodal shape. Such a constraint is to both facilitate the interpretation, and is scientifically plausible, supported by evidences from current literature on human brain development \citep{sowell2002development, sowell2003mapping, wang2012decoding}. Toward that end, we impose 
$$
\bm{\theta}_{kl} \in \mathcal{U}_{kl}, \quad 1 \le k \le l \le K, 
$$
where $\mathcal{U}_{kl}$ complies with one of the following shape constraints, 
\begin{center}
\begin{tabular}{l}
(a) the monotonically increasing shape: \\
\medskip
$\mathcal{U}_{kl} = \{\bm{\theta}_{kl}\in\mathbb{R}^S:\bm{\theta}_{kl,1}\le\ldots\le\bm{\theta}_{kl,s_{kl}}\le\bm{\theta}_{kl,s_{kl}+1}\le\ldots\le\bm{\theta}_{kl,S}\},$ \\ 
(b) the monotonically decreasing shape:  \\
\medskip
$\mathcal{U}_{kl} = \{\bm{\theta}_{kl}\in\mathbb{R}^S:\bm{\theta}_{kl,1}\ge\ldots\ge\bm{\theta}_{kl,s_{kl}}\ge\bm{\theta}_{kl,s_{kl}+1}\ge\ldots\ge\bm{\theta}_{kl,S}\},$ \\
(c) the unimodal shape: \\ 
\medskip
$\mathcal{U}_{kl} = \{\bm{\theta}_{kl}\in\mathbb{R}^S:\bm{\theta}_{kl,1}\le\ldots\le\bm{\theta}_{kl,s_{kl}}\ge\bm{\theta}_{kl,s_{kl}+1}\ge\ldots\ge\bm{\theta}_{kl,S}, \textrm{ for } 1\le s_{kl}\le S\},$ \\
(d) the inverse unimodal shape: \\ 
\medskip
$\mathcal{U}_{kl} = \{\bm{\theta}_{kl}\in\mathbb{R}^S:\bm{\theta}_{kl,1}\ge\ldots\ge\bm{\theta}_{kl,s_{kl}}\le\bm{\theta}_{kl,s_{kl}+1}\le\ldots\le\bm{\theta}_{kl,S}, \textrm{ for } 1\le s_{kl}\le S\}.$ 
\end{tabular}
\end{center}
We briefly note that, imposing a unimodal or inverse unimodal shape constraint consists of two steps: locating the mode, then imposing two monotonic shape constraints. If only imposing the monotonic shape constraint, one simply skips the mode finding step. Figure \ref{fig:c89}, the middle panel, shows the estimated $g^{-1}(\theta_{89})$ under the loss function \eqref{eqn:loss} and the unimodal constraint. It is clearly seen that the connectivity increases with age.

It is also seen from {the middle panel in Figure \ref{fig:c89}} that the estimated connectivity at adjacent time points are often very close, and there are relatively a small number of time points with a substantial change in magnitude. We thus further introduce a fusion type constraint, in addition to the shape constraint, to encourage sparsity in \emph{difference} of the coefficients. Specifically, we consider 
\begin{eqnarray*}
\left \|  \D \btheta_{kl} \right \|_0 \le b_{kl}, \; 1 \le k \le l \le K, \textrm{ where }  
\D =
\begin{pmatrix} 
-1 & 1 & 0 & \cdots &0 & 0 \\
0 & -1 & 1 & \cdots & 0 & 0 \\
\vdots\\
0 & 0 & 0 & \cdots & -1 & 1 \\
\end{pmatrix} \in \mathbb R^{(S-1) \times S},
\end{eqnarray*}
$||\cdot||_0$ is the $\ell_0$ norm and equals the number of non-zero elements, and $b_{kl}$ is the fusion parameter that takes a positive integer value and controls the maximum number of different values in $\btheta_{kl}$. Adding such a fusion constraint again facilitates the model interpretation. Moreover, it is able to reduce the estimation error, as the fusion constraint fits well with the piecewise constant model assumption, and can effectively pool information from adjacent time intervals. Figure \ref{fig:c89}, the right panel, shows the estimated $g^{-1}(\theta_{89})$ under the loss function \eqref{eqn:loss} with both the unimodal constraint and the fusion constraint. It is seen that the connectivity between those two functional modules experiences notable changes when the age is around 9, 13 and 19 years. The resulting estimate is more interpretable than the one without any constraint. We next develop an optimization algorithm to estimate the unknown parameters of our model under the shape and fusion constraints.

\section{Estimation}
\label{sec:estimation}

We aim at the constrained optimization problem, 
\begin{eqnarray}
\label{problem}
\textrm{minimize}_{\bm{\Theta},\bm{\sigma}} \mathcal L(\bm{\Theta},\bm{\sigma}|\c,\bm{A}), \; 
\textrm{ such that } \; \bm{\theta}_{kl} \in \mathcal{U}_{kl}, \textrm{ and } 
\left \|  \D \btheta_{kl} \right \|_0 \le b_{kl}.
\end{eqnarray}
This is a nontrivial optimization as it is neither convex nor biconvex. We propose a sequential estimation procedure with three major steps. We first summarize the procedure in Algorithm~\ref{algo:estimation}, then describe each step in detail. 

In \textbf{step 1}, we seek the minimizer of the loss function \eqref{eqn:loss} without any constraint. We first write \eqref{eqn:loss} as $\mathcal L(\bm{\Theta},\bm{\sigma}|\c,\bm{A})=-\sum_{i=1}^N\log\left\{f_i(\bm A_i | \bm{c}, \bm\Theta_{\tau_i,\cdot},\sigma_{\tau_i})\right\}$, where $\bm\Theta_{s,\cdot}$ is the $s$th row of matrix $\bm\Theta$, $s=1\ldots,S$ and
\begin{equation}
f_i(\bm A_i | \bm{c},\bm\Theta_{\tau_i,\cdot},\sigma_{\tau_i}) = \int_{-\infty}^{+\infty}\prod_{k=1}^K\prod_{l\ge k}^K\frac{\left(e^{\theta_{kl,\tau_i} + \, w_{i}}\right)^{n_{ikl}}}{\left(1+e^{\theta_{kl,\tau_i} + \, w_{i}}\right)^{n_{kl}}} \, \phi(w_{i}, \sigma_{\tau_i}^2) \, \mathrm{d} w_{i}.
\label{fh}
\end{equation}
It involves an integration that cannot be evaluated exactly, but instead can be approximated through the adaptive Gauss-Hermite numerical integration \citep{Pinheiro1995}. That is, the integral $\int f(a)\phi(a)\dd a$ can be approximated as
\begin{eqnarray*}
\int f(a)\phi(a)\dd a \approx\sum_{b=1}^Bf(a_b)u_b,
\end{eqnarray*}
where $B$ is the number of quadrature points used in the approximation, $a_b$ and $u_b$ are the adaptive quadrature nodes and weights \citep{Pinheiro1995}. A reasonable value of $B$ ensures a good approximation of the integral, and we further discuss the choice of $B$ in parameter tuning later. Accordingly, we approximate $f_i(\A_i| \bm{c},\bm\Theta_{\tau_i},\sigma_{\tau_i})$ with
\begin{eqnarray*}
\tilde f_i(\A_i| \bm{c},\bm\Theta_{\tau_i},\sigma_{\tau_i})=\sum_{b=1}^Bu_{ib}\left\{\prod_{k=1}^K\prod_{l\ge k}^K\frac{\left(e^{\theta_{kl,\tau_i}+\sigma_{\tau_h}a_{ib}}\right)^{n_{ikl}}}{\left(1+e^{\theta_{kl,\tau_i}+\sigma_{\tau_i}a_{ib}}\right)^{n_{kl}}}\right\}.
\end{eqnarray*}
We then seek the minimizer, 
\begin{eqnarray} \label{eqn:step1} 
(\tilde{\bm{\Theta}}, \hat{\bm{\sigma}}) = \mathrm{arg\,min}_{\bm{\Theta},\bm{\sigma}} \tilde{\mathcal L}(\bm{\Theta},\bm{\sigma}|\bm{A},\bm{c})=-\sum_{i=1}^N\log\left\{\tilde f_i(\bm A_i | \bm{c},\bm\Theta_{\tau_i},\sigma_{\tau_i})\right\}.
\end{eqnarray}
We achieve this by updating $\bm{\Theta}$ and $\bm{\sigma}$ in an alternating fashion. The block coordinate decent algorithm is used for such an update \citep{tseng2001convergence}.

\begin{algorithm}[t!]
\caption{The sequential optimization procedure of \eqref{problem}.}
\begin{algorithmic}
\STATE \textbf{step 1}: Solve the minimization problem \eqref{eqn:step1} to obtain $\{\tilde{\bm{\Theta}}, \hat{\bm{\sigma}} \}$. 
\REPEAT
\STATE \textbf{step 1.1}: Update $\sigma_s^{(q)}$ given $\{ \bm{\Theta}^{(q-1)},\sigma_1^{(q-1)},\dots,\sigma_{s-1}^{(q-1)},\sigma_{s+1}^{(q-1)},\dots,\sigma_{S}^{(q-1)} \}$, where $\sigma_s^{(q)}$ is the $s$th element of the vector $\bm\sigma^{(q)}$ at iteration $q$, $s=1,\ldots,S$. 
\STATE \textbf{step 1.2}: Update $\bm{\Theta}_{s,\cdot}^{(q)}$ given $\{ \bm{\Theta}_{1,\cdot}^{(q-1)}, \ldots, \bm{\Theta}_{s-1,\cdot}^{(q-1)}, \bm{\Theta}_{s+1,\cdot}^{(q-1)}, \ldots, \bm{\Theta}_{S,\cdot}^{(q-1)}, \bm\sigma^{(q)}\}$, where $\bm{\Theta}_{s,\cdot}^{(q)}$ is the $s$th row of the matrix $\bm\Theta^{(q)}$ at iteration $q$, $s=1,\ldots,S$.
\STATE \textbf{step 1.3}: Update the iteration number $q = q + 1$.
\UNTIL{the objective function $\tilde{\mathcal L}(\bm{\Theta},\bm{\sigma}|\bm{A},\bm{c})$ in \eqref{eqn:step1} converges.}
\STATE \textbf{step 2}: Solve the shape constrained problem \eqref{eqn:step2} to obtain $\check{\bm{\Theta}}$.
\STATE \textbf{step 3}: Solve the fusion constrained problem \eqref{eqn:step3} to obtain $\hat{\bm{\Theta}}$.
\end{algorithmic}\label{algo:estimation}
\end{algorithm}

In \textbf{step 2}, we add the shape constraint to the loss function in \eqref{eqn:loss}, by computing a projection of $\tilde\btheta_{kl}$, the column of $\tilde{\bm{\Theta}}$, to the set of shape constrained sequences,
\begin{eqnarray} \label{eqn:step2}
\check\btheta_{kl}=\mathrm{arg\,min}_{v_1,\ldots,v_S} \sum_{s=1}^S(v_s-\tilde\theta_{kl,s})^2, \textrm{ such that } (v_1,\ldots,v_S) \in\mathcal{U}_{kl}.
\end{eqnarray}
That is, we find the vector $\check\btheta_{kl} = (\check\theta_{kl,1}, \ldots, \check\theta_{kl,S})^\top$ such that it satisfies the shape constraint specified in $\mathcal{U}_{kl}$, meanwhile minimizing the sum of squared error with respect to $\tilde\btheta_{kl} = (\tilde\theta_{kl,1}, \ldots, \tilde\theta_{kl,S})^\top$. For the unimodal shape constraint, we first find the mode of the sequence, by performing an isotonic regression for each possible candidate, then taking the one that yields the smallest sum of squared error as the estimated mode \citep{turner1997locating}. Once we locate the mode, finding a unimodal sequence breaks into finding two monotonic sequences, one monotonically increasing before the mode and the other monotonically decreasing after the mode, and each can be achieved by an isotonic regression. The isotonic regression is carried out using an efficient linear algorithm, the pool adjacent violators algorithm \citep{barlow1972statistical}. For the inverse unimodal constraint, the estimation can be carried out similarly. For the monotonic shape constraint, we simply skip the mode finding step, and the rest of implementation is the same.  

In \textbf{step 3}, we further add the fusion constraint to the shape constrained estimate $\check\btheta_{kl}$, by seeking 
\begin{eqnarray} \label{eqn:step3} 
\hat\btheta_{kl}=\mathrm{arg\,min}_{v_1,\ldots,v_S} \sum_{s=1}^S(v_s-\check\theta_{kl,s})^2, \textrm{ such that } \|\D \bv\|_0 \le b_{kl},
\end{eqnarray}
where $\bv = (v_1,\ldots,v_S)^\top$, and $b_{kl}$ is the fusion tuning parameter. A common way to incorporate the fusion penalty is through soft thresholding, such as in fused lasso \citep{tibshirani2005sparsity}, though it would yield a biased estimator \citep{rinaldo2009properties}. Instead, we employ a hard thresholding approach. Specifically, for a vector $\bv = (v_1, \ldots, v_S)^\top \in \mathbb R^S$ and a positive integer number $b \le S$, we first truncate the vector $\D \bv$, such that we keep the largest $(b-1)$ entries in their absolute values, and set the rest to zero. This would in effect yield that there are $b$ groups of distinct values in $\bv$. We then average the values of $\bv$ within each of those $b$ groups. We call this a fusion operator, and denote it by $\mathrm{Fuse}(\bv, b)$. As a simple illustration, let $\bv = (0.1, 0.2, 0.3, 0.5, 0.6)^\top$ and $b=2$. Then $\D \bv=(0.1,0.1,0.2,0.1)^\top$ and after truncation, we obtain $(0,0,0.2,0)^\top$. This in effect suggests that there are two groups in $\bv$, with the first three entries $v_1,v_2,v_3$ of $\bv$ belong to one group, and the last two entries $v_4,v_5$ belong to the other. We then average the values of $\bv$ in each group, and obtain that $\mathrm{Fuse}(\bv, b) = (0.2, 0.2, 0.2, 0.55, 0.55)^\top$.  

It is important to note that, the added fusion structure in the third step actually \emph{preserves} the shape constraint from the second step, in that the estimator after applying the fusion constraint still belongs to the parameter space under the shape constraint. The next lemma characterizes this result, and its proof is given in the Supplementary Materials. 
 
\begin{lemma}
Let $\mathcal{U}$ comply with one of the following shape constraints on sequences $\bv \in \mathbb R^{S}$: monotonically increasing, monotonically decreasing, unimodal, or inverse unimodal. For any $\bv\in\mathcal{U}$ and $b\in\mathbb{N}^+$, we have $\textrm{Fuse}(\bv, b)\in \mathcal{U}$.
\label{lemma_fuse}
\end{lemma}

Our proposed three-step estimation procedure involves a number of tuning parameters, including the number of intervals $S$, the number of quadrature points $B$ in the integral approximation, and the fusion penalty parameters $b_{kl}$. For $S$, in our numerical analysis, we have experimented with a number of different values and found our final estimate is not overly sensitive to the choice of $S$. See Sections \ref{sec:simulation} and \ref{sec:realdata} for more details. For $B$, we have also experimented with a range of values {between 1 and 25}, which is often deemed sufficient to provide a good approximation to the numerical integration. The final estimates with $B$ in this range are very close. As a greater $B$ value implies a higher computational cost, we fix $B = 5$ in our subsequent analysis. For $b_{kl}$, we propose to use the following BIC criterion for tuning, 
\begin{eqnarray*}
\mathrm{BIC} = -2 \tilde{\mathcal L}_{kl}(\hat\btheta_{kl},\bm{\hat\sigma}|\bm{A},\bm{c})+\log \, N \times df(\hat\btheta_{kl}).
\end{eqnarray*}
In the above BIC criterion, we have
{
$$
\tilde{\mathcal L}_{kl}(\hat\btheta_{kl},\bm{\hat\sigma}|\bm{A},\bm{c})=\sum_{i=1}^{N}\log \left\{\sum_{b=1}^B u_{ib}\frac{\left(e^{\theta_{kl,\tau_i} + \, \sqrt{2} \sigma_{\tau_i} a_{ib}}\right)^{n_{ikl}}}{\left(1+e^{\theta_{kl,\tau_i} + \, \sqrt{2} \sigma_{\tau_i} a_{ib}}\right)^{n_{kl}}}\right\}
$$ and} $df(\hat\btheta_{kl})$ denotes the degrees of freedom, which is defined as the number of unique elements in the vector $\hat\btheta_{kl}$.

\section{Theory}
\label{sec:theory}

We next investigate the asymptotic property of our proposed estimator as the sample size $N$ tends to infinity. We first introduce some notations. For a function $v(t) \in L^2([0,1])$, we define its function $\ell_q$ norm as $\|v\|_q=\left( \int_0^1|v(t)|^q dt \right)^{1/q}$. 
For two sequences $\{a_n\}$ and $\{b_n\}$, we write $a_n\prec b_n$ if $a_n/b_n\rightarrow 0$ as $n\rightarrow\infty$. We write $a_n\asymp b_n$ if $a_n\le cb_n$ and $a_n\ge c'b_n$ for some positive constants $c$ and $c'$.
For $1\le k\le l\le K$, let $\theta_{kl}(t)$ denote the true underlying function, which is assumed to be piecewise constant and comply with one of the following shape constraints: monotonically increasing, monotonically decreasing, unimodal or inverse unimodal. Let $p_{kl}$ denote the true number of constant intervals of $\theta_{kl}(t)$, and $p = \max_{1 \le k, l \le K} p_{kl}$. We write $(0,1]=\cup_{j=1}^{p_{kl}}(t_{kl,j-1},t_{kl,j}]$, for $0=t_{kl,0}<t_{kl,1}<\ldots<t_{kl,p_{kl}}=1$, and 
\[
\theta_{kl}(t)=\sum_{j=1}^{p_{kl}} \theta_{kl,j} I\left\{t \in (t_{kl, j-1},t_{kl,j}]\right\}.
\]
We assume that $|t_{kl, j-1}-t_{kl,j}|\asymp \frac{1}{p_{kl}}$ for $j=1,\ldots,p_{kl}$. Given a chosen number of intervals $S$, the vector estimates $\check{\btheta}_{kl}$ from \eqref{eqn:step2} and $\hat{\btheta}_{kl}$ from \eqref{eqn:step3}, we define the corresponding function estimates as 
\begin{eqnarray*}
\check\theta_{kl}(t) & = & \sum_{s=1}^{S} \check\theta_{kl,s} I\left\{t \in \left( \frac{s-1}{S}, \frac{s}{S} \right]\right\}, \\
\hat\theta_{kl}(t) & = & \sum_{s=1}^{S} \hat\theta_{kl,s} I\left\{t \in \left( \frac{s-1}{S}, \frac{s}{S} \right]\right\}. 
\end{eqnarray*}

We next introduce a set of main regularity conditions, and give some additional technical conditions in the Supplementary Materials. 
\begin{assumption}
\label{ass:size} 
Assume the number of intervals $S$ and the number of subjects $N$ satisfy that $S=O(N^{1-\alpha})$ for some $0 < \alpha < 1$.
\end{assumption}
\noindent 
This assumption allows the number of intervals $S$ to diverge along with the subject size $N$. The condition $\alpha <1$ ensures that there are enough number of samples in each interval to guarantee the performance of our estimation in the first step.

\begin{assumption}
\label{ass:tuning}
Assume that $p\prec S^{1/3}$ and the tuning parameter $b_{kl} \ge 2p$.
\end{assumption}
\noindent 
This assumption allows the true number of constant intervals in the true underlying function to diverge. Moreover, by placing a lower bound on the tuning parameter $b_{kl}$, the fusion step would not incorrectly merge two distinctive groups of coefficients. This is similar to the condition required in the hard thresholding operator in sparse learning problems \citep{yuan2013truncated}, and is also analogous to the condition on the tuning parameter in soft thresholding based fused lasso \citep{rinaldo2009properties}. 

\begin{assumption}
\label{ass:gap} 
Denote the gaps in $\theta_{kl}(t)$ as $\Delta_{kl,j}=|\theta_{kl,j}-\theta_{kl,j-1}|$, $j=1,\ldots,p_{kl}$. Assume that $p\sqrt{(p \, \log \, S) / N} \prec \Delta_{kl,j} \prec \sqrt{(S \, \log \, S) / N}$, $1 \leq k,l \leq K, j=1,\ldots,p_{kl}$.
\end{assumption}
\noindent 
This assumption places both a lower bound and an upper bound on the gaps among sequential distinct values of $\theta_{kl}(t)$. The condition on the lower bound is analogous to the minimal signal condition of sparse coefficients in high-dimensional regressions \citep{fan2014strong}. The condition on the upper bound indicates that, as the number of intervals in $\theta_{kl}$ increases, the signal in $\theta_{kl}$ grows at a bounded rate. This is needed to control the approximation error. Moreover, the technical condition (C4) in the Supplementary Materials ensures the identifiably of the unique minimizer. 

Next we present the main theorem of our proposed estimator.
\begin{theorem}
Under Assumptions \ref{ass:size} - \ref{ass:gap} and the technical condition (C4) in the Supplementary Materials, we have 
\begin{eqnarray} \label{eqn:smallerMSE}
\| \check\theta_{kl}(t) - \theta_{kl}(t) \|_2 \ge \| \hat\theta_{kl}(t) - \theta_{kl}(t) \|_2, 
\end{eqnarray}
where the equality holds if and only if $\check\theta_{kl}(t) = \hat\theta_{kl}(t)$. Furthermore, we have   
\begin{eqnarray} \label{eqn:errorbound}
\| \hat\theta_{kl}(t) - \theta_{kl}(t) \|^2_2 \le C\left(\frac{p \, \log \,S}{N}\right), 
\end{eqnarray}
for some positive constant $C$, and $\| \cdot \|_2$ denotes the function $\ell_2$ norm.

\label{thm1}
\end{theorem}

We give an outline of the proof here, and present the complete proof in the Supplementary Materials. The fact that the true underlying connectivity function $\theta_{kl}(t)$ is a step function with unknown time intervals, and that the estimation is multi-step, have posed challenges in obtaining the asymptotic property of our estimator. Based on \cite{domowitz1982misspecified}, we first show that $\tilde\btheta_{kl}$, which is obtained from \eqref{eqn:step1} in Step 1, is asymptotically normal and consistent. It is also noteworthy that, the estimation of $\tilde\btheta_{kl}$ can be formulated as a generalized mixed partial linear regression problem \citep{heckman1986spline,chen2015semiparametric,lian2015variance}. As such, $\tilde\btheta_{kl}$ can be viewed as the estimator when the unknown function in this model takes a step function form. The asymptotic property of this type of estimator has not yet been studies in the literature. Next we follow \cite{chatterjee2015adaptive}, which establishes the adaptive risk of the unimodal least square estimator, to obtain the asymptotic property of the shape constrained estimator $\check\btheta_{kl}$ from Step 2. Finally, we show that the error of the fused and  shape constrained estimator is no greater than that of the shape constrained estimator using the property of our hard thresholding fusion operator. This is reflected in the inequality in \eqref{eqn:smallerMSE}, which offers a theoretical justification of the fusion step in our procedure. We then establish the error bound for the final estimator $\hat\theta_{kl}(t)$ from Step 3. 

We also make some remarks about the derived error bound. First, when the true function $\theta_{kl}(t)$ is indeed piecewise constant, the error bound obtained in \eqref{eqn:errorbound} for our estimator $\hat\theta_{kl}(t)$ is analogous to the bound that is commonly seen in sparse regressions, which is in the form of $\mathrm{sparsity} \times \log(\mathrm{dimension})/(\mathrm{sample \ size})$ \citep{Negahban2012}. In our setup, $p$ plays the role of the sparsity quantity, and $S$ the role of the dimension. Second, when the true function $\theta_{kl}(t)$ is not piecewise constant, we work under a misspecified model. Since the number of intervals $S$ is allowed to grow with the sample size $N$, the difference between the true function and the piecewise constant approximation tends to zero as $S$ tends to infinity. In this case, the number of distinct values in the approximation function scales with $S$, and the corresponding error bound becomes $C(S \, \log \, S) / N$. This is analogous to the error bound that is commonly seen in non-sparse regressions, which is in the form of $\mathrm{dimension} \times \log(\mathrm{dimension}) / (\mathrm{sample \ size})$ {\citep{buhlmann2011statistics}}.

\section{Simulations}
\label{sec:simulation}

Next we investigate the finite sample performance of our proposed method. We generated $N$ subjects with the age variable simulated from a uniform distribution on $[0,1]$. For each subject a network was generated based on a stochastic blockmodel, with $n$ nodes belonging to $K$ communities. We fix $n=75$, $K=2$, and the two communities have numbers of nodes equal to $n_1 = 50$ and $n_2 = 25$, respectively. 

We consider different ways of generating the within-community and between-community  connectivity functions $\{\theta_{11}(t), \theta_{12}(t), \theta_{22}(t)\}$ and the variance function of the random-effect $\sigma^2(t)$. 

\begin{figure}[t!]
\centering
\includegraphics[scale=0.405, trim=10mm 10mm 5mm 0 ]{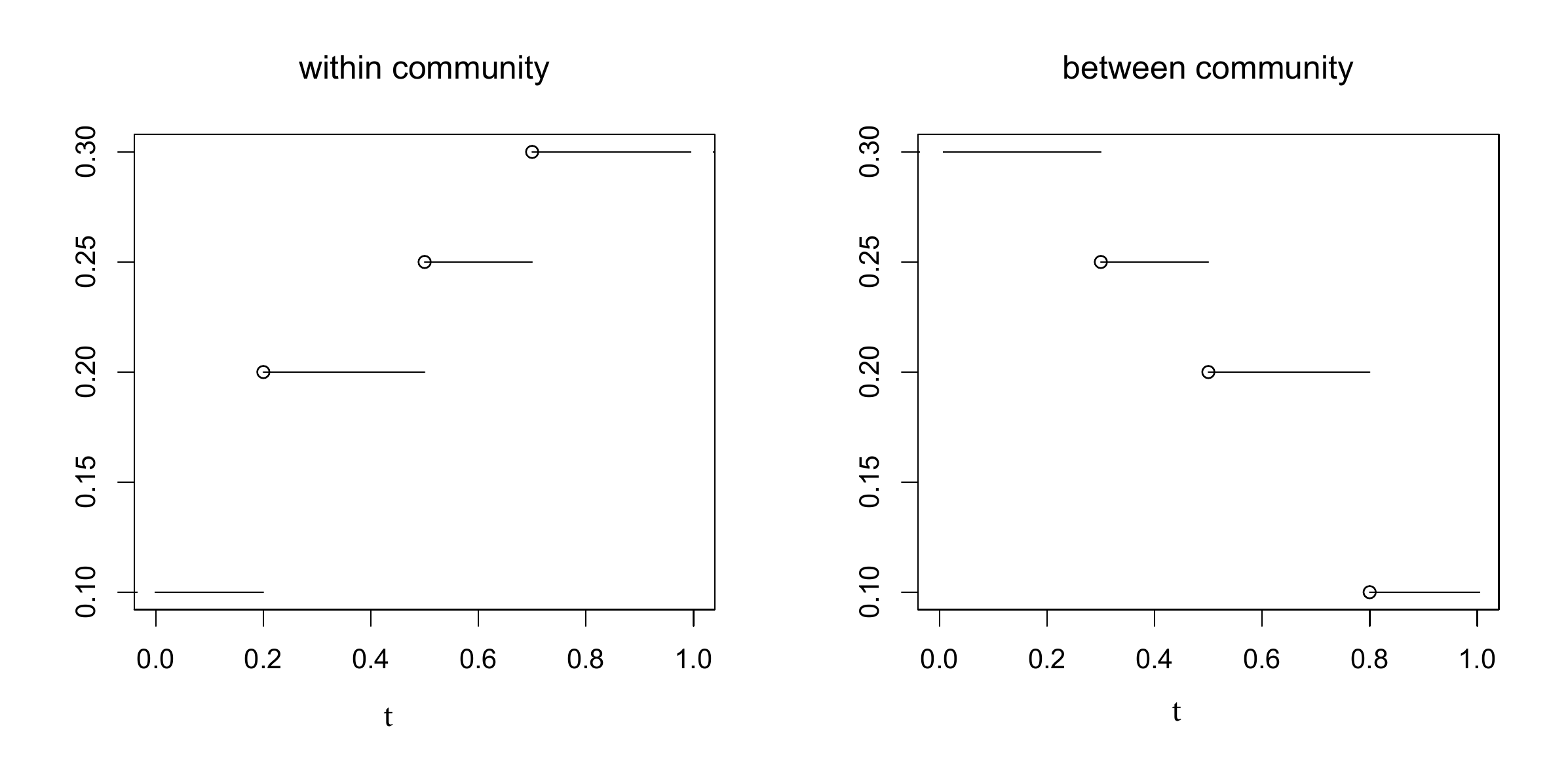} \\
\includegraphics[scale=0.39, trim=10mm 15mm 5mm 0 ]{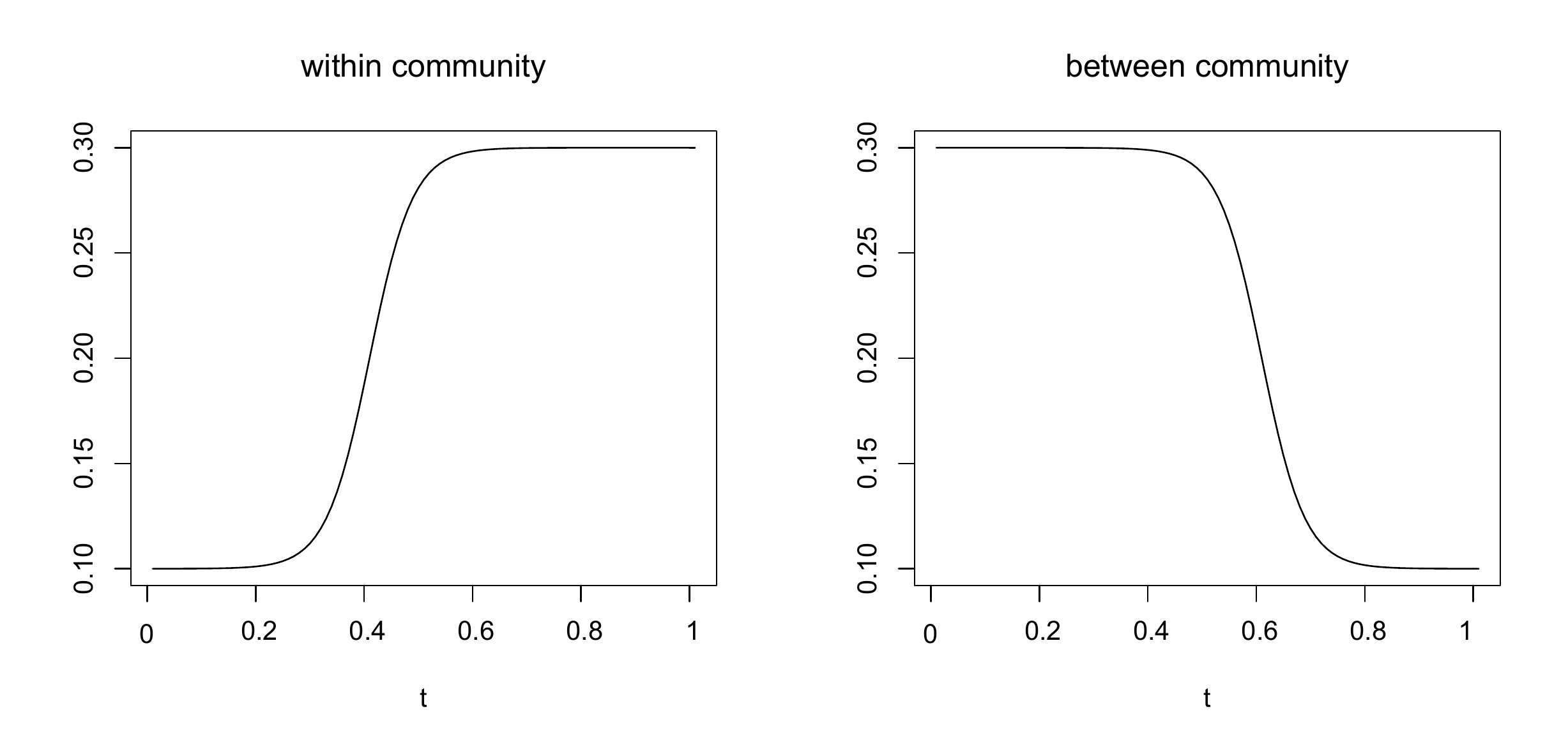} 
\caption{Two scenarios of the true within-community and between-community connecting probabilities as functions of time.}
\label{fig:connectivity}
\end{figure}

\smallskip
\textbf{Example A}: The connectivity functions are time-varying, while the variance of the random-effect is fixed over time. Specifically,  we set $\logit^{-1}\{\theta_{11}(t)\}$ and $\logit^{-1}\{\theta_{22}(t)\}$ to take the functional form as shown in the upper left panel and $\logit^{-1}\{\theta_{12}(t)\}$ as shown in the upper right panel of Figure~\ref{fig:connectivity}. In this setting, the within-community and between-community time-varying connecting probability follow different trajectories. The within-community connecting probability is monotonically increasing with $[0,0.2]$, $(0.2,0.5]$, $(0.5,0.7]$, and $(0.7,1]$ as true intervals, whereas the between-community connecting probability is monotonically decreasing with $[0,0.3]$, $(0.3,0.5]$, $(0.5,0.8]$, $(0.8,1]$ as the true intervals. We set $\sigma = 0.1$. 

\smallskip
\textbf{Example B}: Both the connectivity functions and the variance function are time-varying. Specifically, we choose the same $\{\theta_{11}(t), \theta_{12}(t), \theta_{22}(t)\}$ as in Example A, while we set $\sigma(t)=0.2-0.1I(t>0.5)$. This function is motivated by the scientific finding that there is usually a higher variability in functional connectivity when subjects are at a younger age \citep{nomi2017moment}. 

\smallskip
\textbf{Example C}: Both the connectivity functions and the variance function are time-varying. Moreover, the connectivity functions are no longer piecewise constant, but instead continuous functions. This example thus evaluates the performance of our method under model misspecification. Specifically, we set $\logit^{-1}\{\theta_{11}(t)\}$ and $\logit^{-1}\{\theta_{22}(t)\}$ to take the functional form as shown in the lower left panel and $\logit^{-1}\{\theta_{12}(t)\}$ as shown in the lower right panel of Figure~\ref{fig:connectivity}. In this setting, the within-community connectivity is monotonically increasing and goes through a continuous and notable increase between $t=0.2$ and $t=0.6$; the between-community connectivity is monotonically decreasing and goes through a continuous decrease between $t=0.4$ and $t=0.8$. We again set $\sigma(t)=0.2-0.1I(t>0.5)$.

We compare three estimation methods. The first is our proposed mixed-effect time-varying stochastic blockmodel estimator with shape and fusion constraints. The second is the unconstrained estimator obtained from \eqref{eqn:step1}. The third is to fit a cubic B-spline with nine interior knots equally spaced in $[0,1]$. For the first two methods, we divide $[0,1]$ into $S$ equal-length intervals, and we vary the value of $S$. We impose the unimodal shape constraint and use BIC to tune the fusion penalty parameters. The estimation error is evaluated based on the criterion, $\int_{0}^1 \{\theta(t)-\hat\theta(t)\}^2\dd t \, / \int_{0}^1\theta(t)^2\dd t$, where $\hat\theta(t)$ denotes the estimated function of the truth $\theta(t)$.

\begin{figure}[t!]
 \includegraphics[scale=0.425,right]{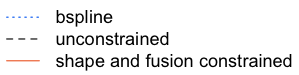}
\centering
\includegraphics[scale=0.425, trim=10mm 10mm 5mm 10mm]{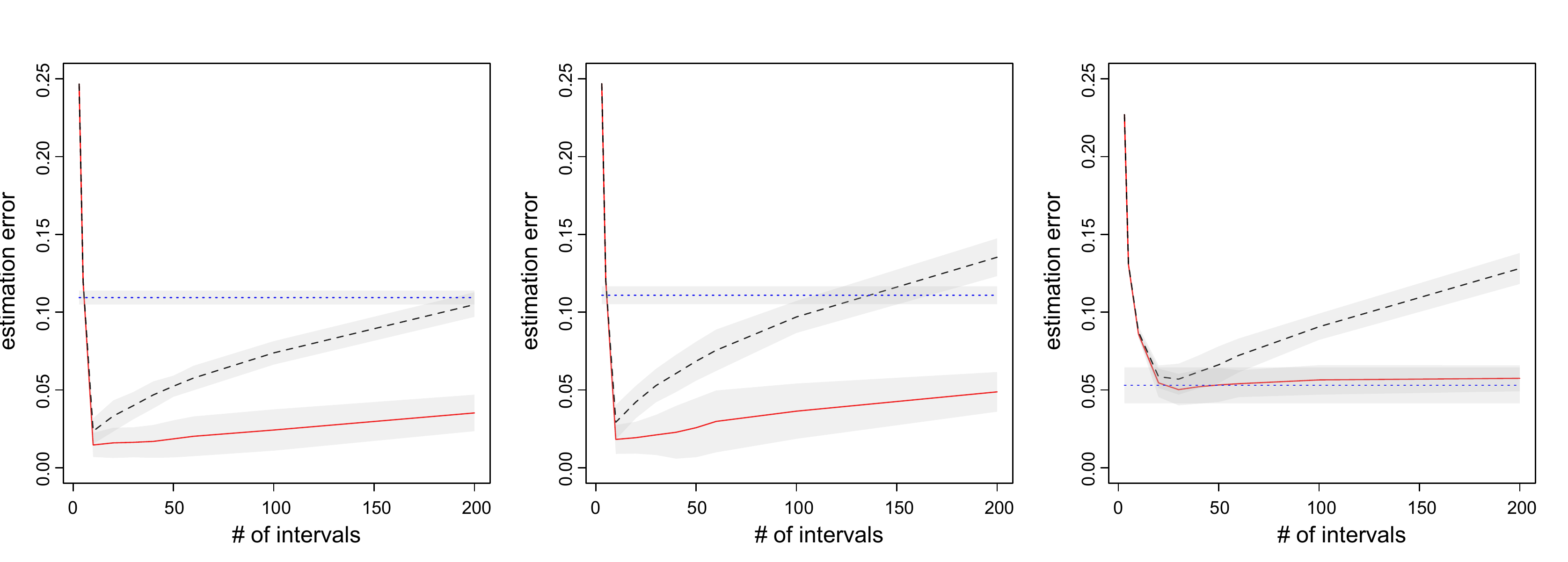}
\caption{Estimation error of three different methods: the proposed shape and fusion constrained estimator (solid line), the unconstrained estimator (dashed line), and the cubic B-spline estimator (dotted line). The grey bands are the 95\% confidence interval.}
\label{fig:sim}
\end{figure}

Figure~\ref{fig:sim} reports the results based on 50 data replications, while the sample size is fixed at $N = 600$. For Example A (the left panel), we observe that our proposed fused and shape constrained estimator achieves the best overall performance, as long as the number of intervals $S$ is reasonably large ($S \ge 10$). Moreover, its performance is stable for a wide range of values of $S$. This is an appealing feature of our estimator. By contrast, the unconstrained estimator is much more sensitive to the choice of $S$. This is because it does not borrow information across different time intervals as the constrained estimator does. The spline estimator also performs poorly, since the underlying true function is piecewise constant rather than smooth. For Example B (the middle panel), we again observe that our proposed estimator performs the best, and remains relatively stable across a wide range of values of $S$. By comparison, the unconstrained estimator performs worse when the random-effect variance varies over time. The spline estimator continues to perform poorly in this setting. For Example C (the right panel), we see that our proposed method maintains a competitive performance, even though the underlying true functions are continuous and our model is misspecified. This is due to the property that our estimator effectively pools information across different time intervals and across subjects. By contrast, the unconstrained estimator performs poorly. This setting favors the spline estimator; however, our method performs about the same as the spline solution. 

\begin{table}
\setlength{\tabcolsep}{5pt} 
\fbox{%
\begin{tabular}{c|c|c|c|c|c|c|c}
\multirow{2}{*}{Method}                                                                               & \multirow{2}{*}{$N$} & \multicolumn{2}{c|}{Example A}                                                                                        & \multicolumn{2}{c|}{Example B}                                                                                        & \multicolumn{2}{c}{Example C}                                                                                        \\ \cline{3-8} 
                                                                                                      &                      & $S=20$                                                    & $S=50$                                                    & $S=20$                                                    & $S=50$                                                    & $S=20$                                                    & $S=50$                                                    \\ \hline
\multirow{2}{*}{\begin{tabular}[c]{@{}c@{}}shape and fusion \\ constrained \\ estimator\end{tabular}} & 200                  & \begin{tabular}[c]{@{}c@{}}0.0335\\ (0.0093)\end{tabular} & \begin{tabular}[c]{@{}c@{}}0.0545\\ (0.0115)\end{tabular} & \begin{tabular}[c]{@{}c@{}}0.0490\\ (0.0148)\end{tabular} & \begin{tabular}[c]{@{}c@{}}0.0719\\ (0.0144)\end{tabular} & \begin{tabular}[c]{@{}c@{}}0.0720\\ (0.0078\end{tabular}  & \begin{tabular}[c]{@{}c@{}}0.0762\\ (0.0077)\end{tabular} \\ \cline{2-8} 
                                                                                                      & 600                  & \begin{tabular}[c]{@{}c@{}}0.0171\\ (0.0051)\end{tabular} & \begin{tabular}[c]{@{}c@{}}0.0197\\ (0.0056)\end{tabular} & \begin{tabular}[c]{@{}c@{}}0.0191\\ (0.0051)\end{tabular} & \begin{tabular}[c]{@{}c@{}}0.0236\\ (0.0090)\end{tabular} & \begin{tabular}[c]{@{}c@{}}0.0551\\ (0.0041)\end{tabular} & \begin{tabular}[c]{@{}c@{}}0.0541\\ (0.0073)\end{tabular} \\ \hline
\multirow{2}{*}{\begin{tabular}[c]{@{}c@{}}unconstrained\\ estimator\end{tabular}}                    & 200                  & \begin{tabular}[c]{@{}c@{}}0.0596\\ (0.0063)\end{tabular} & \begin{tabular}[c]{@{}c@{}}0.0938\\ (0.0068)\end{tabular} & \begin{tabular}[c]{@{}c@{}}0.0784\\ (0.0117)\end{tabular} & \begin{tabular}[c]{@{}c@{}}0.1195\\ (0.0105)\end{tabular} & \begin{tabular}[c]{@{}c@{}}0.0811\\ (0.0068)\end{tabular} & \begin{tabular}[c]{@{}c@{}}0.1124\\ (0.009)\end{tabular}  \\ \cline{2-8} 
                                                                                                      & 600                  & \begin{tabular}[c]{@{}c@{}}0.0337\\ (0.0040)\end{tabular} & \begin{tabular}[c]{@{}c@{}}0.0529\\ (0.0041)\end{tabular} & \begin{tabular}[c]{@{}c@{}}0.0409\\ (0.0050)\end{tabular} & \begin{tabular}[c]{@{}c@{}}0.0665\\ (0.0058)\end{tabular} & \begin{tabular}[c]{@{}c@{}}0.0586\\ (0.0039)\end{tabular} & \begin{tabular}[c]{@{}c@{}}0.0674\\ (0.0065)\end{tabular} \\ \hline
\multirow{2}{*}{\begin{tabular}[c]{@{}c@{}}cubic B-spline \\ estimator\end{tabular}}                  & 200                  & \multicolumn{2}{c|}{\begin{tabular}[c]{@{}c@{}}0.1156\\ (0.0043)\end{tabular}}                                        & \multicolumn{2}{c|}{\begin{tabular}[c]{@{}c@{}}0.1216\\ (0.0065)\end{tabular}}                                        & \multicolumn{2}{c}{\begin{tabular}[c]{@{}c@{}}0.0679\\ (0.0088)\end{tabular}}                                        \\ \cline{2-8} 
                                                                                                      & 600                  & \multicolumn{2}{c|}{\begin{tabular}[c]{@{}c@{}}0.1087\\ (0.0025)\end{tabular}}                                        & \multicolumn{2}{c|}{\begin{tabular}[c]{@{}c@{}}0.1108\\ (0.0028)\end{tabular}}                                        & \multicolumn{2}{c}{\begin{tabular}[c]{@{}c@{}}0.0533\\ (0.0052)\end{tabular}}                                        \\ 
\end{tabular}}
\caption{\label{tab:sim} Average estimation errors (with standard deviations in the parenthesis) of three methods with varying sample size $N$ and number of intervals $S$.}
\end{table}

We also vary the sample size $N=200,600$ for different values of $S=20,50$ {in the above three examples}. Table~\ref{tab:sim} reports the average estimation error and the standard deviation (in parenthesis) for the three methods based on 50 data replications. It is clearly seen that the performance of the proposed estimator improves with a larger sample size $N$ and a smaller number of intervals $S$. This observation agrees with our theoretical results. Consistent with the pattern shown in Figure~\ref{fig:sim}, we see that our proposed estimator achieves the best performance in all combinations of $N$ and $S$ in Examples A and B. In Example C, our proposed estimator performs similarly as the cubic B-spline, and both outperform the unconstrained one.

\section{Brain development study}
\label{sec:realdata}
We revisit the motivating example of brain development study in youth that is introduced in Section \ref{sec:introduction}. We analyzed a resting-state fMRI dataset from the ADHD-200 Global Competition (\url{http://fcon_1000.projects.nitrc.org/indi/adhd200/}). Resting-state fMRI characterizes functional architecture and synchronization of brain systems, by measuring spontaneous low-frequency blood oxygen level dependent signal fluctuations in functionally-related brain regions. During fMRI image acquisition, all subjects were asked to stay awake and not to think about anything under a black screen. All fMRI scans have been preprocessed, including slice timing correction, motion correction, spatial smoothing, denoising by regressing out motion parameters and white matter and cerebrospinal fluid time, and band-pass filtering. Each fMRI image was summarized in the form of a network, with the nodes corresponding to 264 seed regions of interest in the brain, following the cortical parcellation system defined by \citet{Power2011}, and the links recording partial correlations between pairs of those 264 nodes. Moreover, those 264 nodes have been partitioned into 10 functional modules corresponding to the major resting-state networks defined by \citet{Smith2009}. These modules are \emph{(1) medial visual, (2) occipital pole visual, (3) lateral visual, (4) default mode, (5) cerebellum, (6) sensorimotor, (7) auditory, (8) executive control, (9) frontoparietal right}, and \emph{(10) frontoparietal left}, and they form the $K=10$ communities in our stochastic blockmodeling. The original data included both combined types of attention deficit hyperactivity disorder subjects and typically developing controls. We focused on the control subjects only in our analysis, as our goal is to understand brain development in healthy children and adolescents. This results in $N = 491$ subjects. Their age ranges between 7 and 20 years, and is continuously measured.

\begin{figure}[t!]
\includegraphics[scale=0.45,right,trim=0 0 5mm 0]{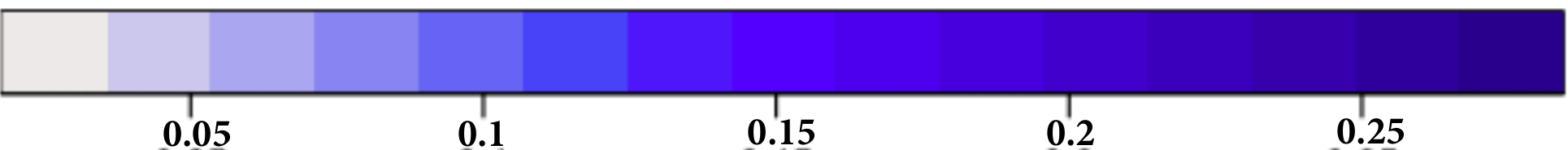}
\centering
\includegraphics[scale=0.425, trim=4mm 0 0 0]{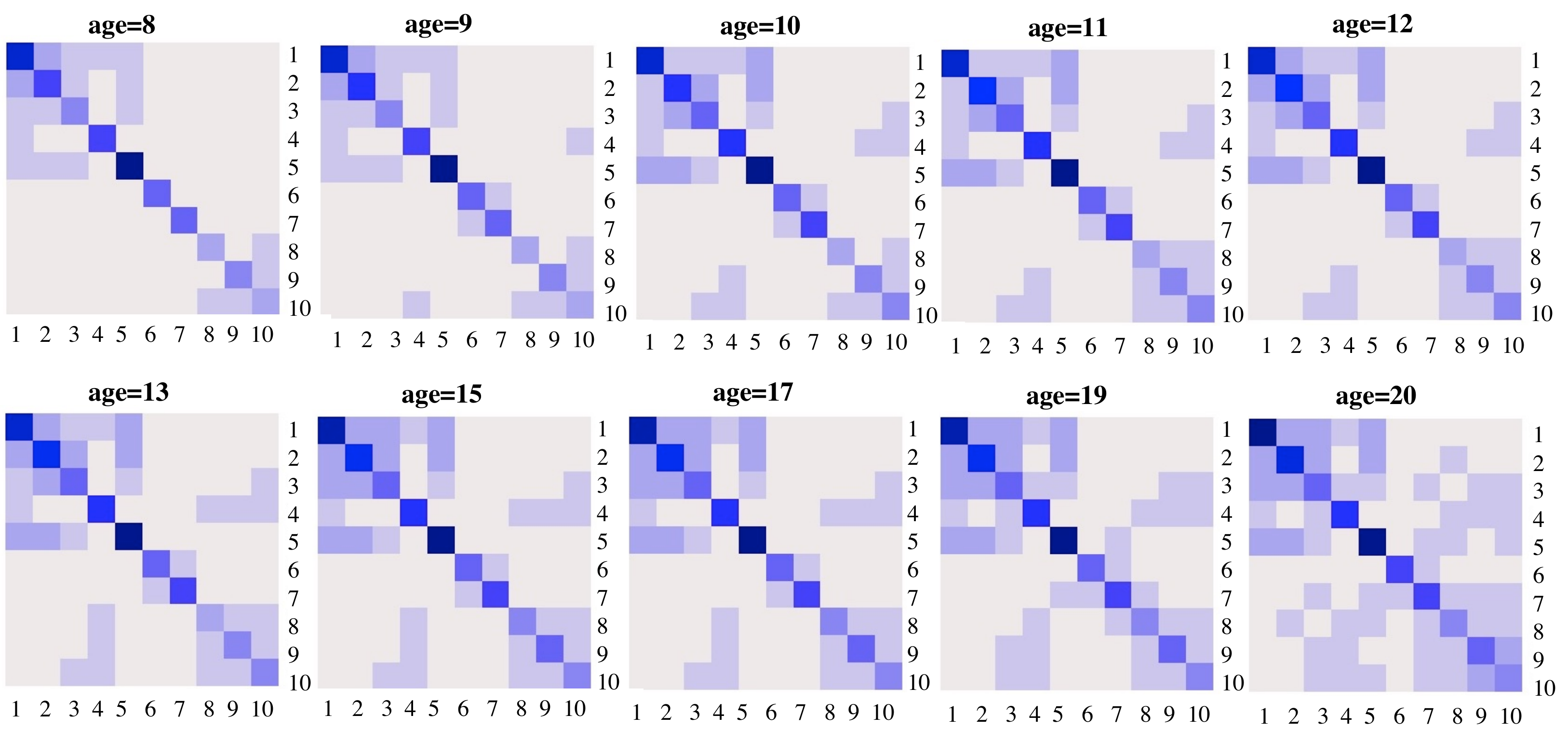}
\caption{Heat maps of the estimated within- and between-communities connecting probabilities at different ages.}
\label{fig:all}
\end{figure}

We applied our method to this data. We partitioned the time interval such that each interval contains about the same number of subjects $20$. We have also experimented with other choices of $S$, and obtained similar results. We further discuss the sensitivity of our result as a function of $S$ later. We applied both the unimodal and inverse unimodal shape constraints, and selected the one with a smaller estimation error. We used BIC to select the fusion tuning parameter. Figure \ref{fig:all} reports the heat maps of the connecting probabilities both within and between the ten functional modules when age takes integer values from 8 to 20.  In principle we can estimate the connecting probabilities for any continuous-valued age. 

We first make some observations of the overall patterns of the connectivity as a function of age. It is seen that the within-community connectivity is greater than the between-community connectivity, by noting generally much darker cells on the diagonals. This observation agrees with the literature that brain regions within the same functional module share high within-module connectivity \citep{Smith2009}. It is also seen that the between-community connectivity grows stronger with age, by noting an increasing number of dark-colored off-diagonal cells as age increases. 

We next examine the between-community connectivity patterns. It is seen from Figure \ref{fig:all} that the connectivity between the 4th community, \emph{default mode network}, and other modules increases with age. This observation supports the existing literature that the default mode module has increasingly synchronized connections to other modules \citep{grayson2017development}. It is also observed that there is increased connectivity between the 5th community, \emph{cerebellum}, and other modules, which agrees with a similar finding in \cite{fair2009functional}. Moreover, it is seen that the 6th community, \emph{sensorimotor}, is segregated from all other communities, with a low connecting probability between this community and the rest. A similar finding has been reported in \citet{grayson2017development}. In addition, we have noted that the first three communities, all involved with visual function, exhibit a high between-community connectivity even at young ages, and this connectivity strengthens with age; {the \emph{frontoparietal} modules (Communities 9 and 10) show increased connectivity to the \emph{executive control} module (Community 8) with age}. Such results have not been reported in the literature and are worth further investigation.

\begin{figure}[t!]
\centering
\includegraphics[scale=0.65, trim=10mm 10mm 0 15mm]{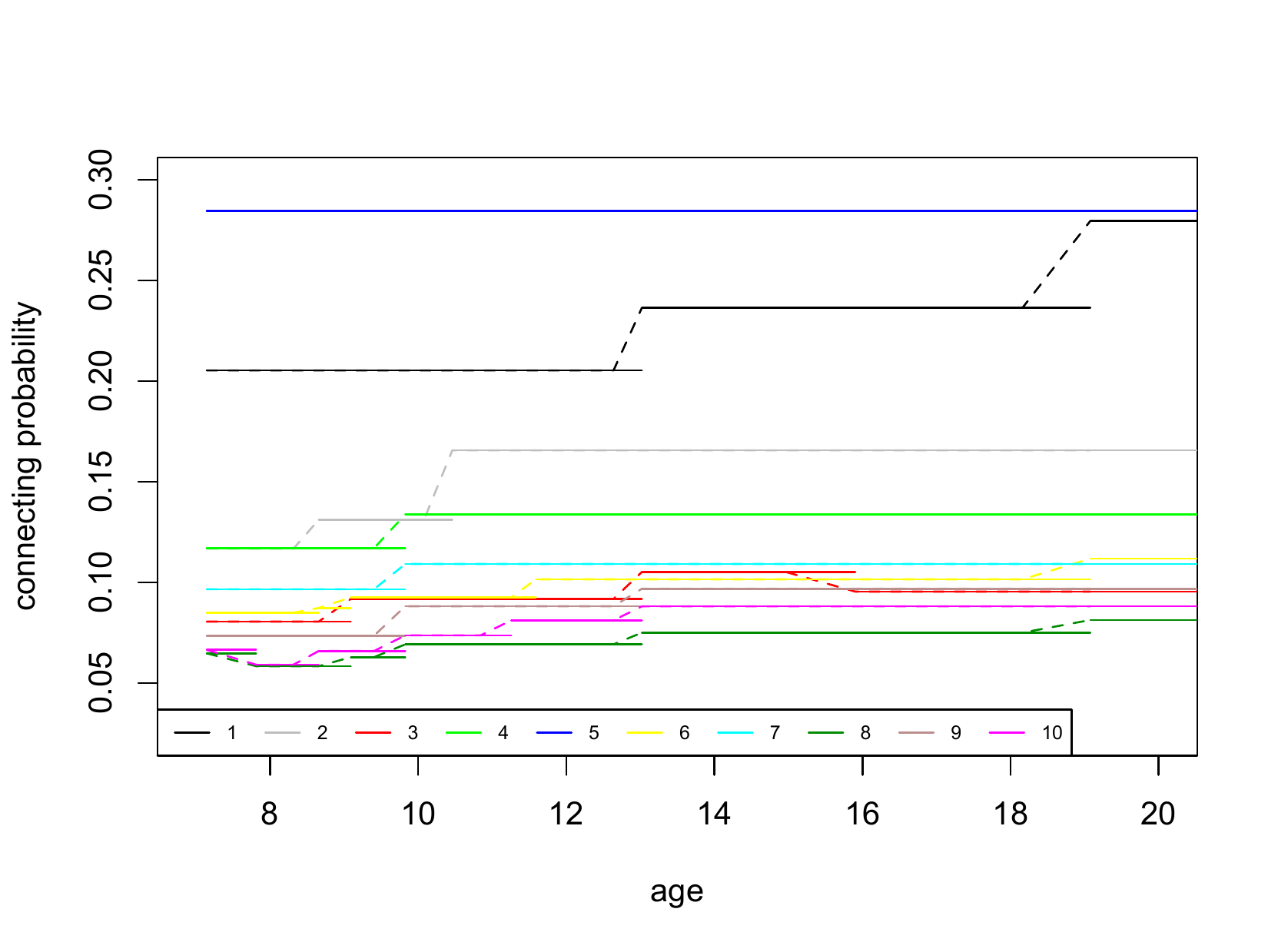}
\caption{The estimated within-community connecting probabilities as functions of age.}
\label{fig:within}
\end{figure}

We then examine the within-community connectivity patterns, which are shown in Figure \ref{fig:within}. It is seen that the 5th community, \emph{cerebellum}, has a high within-community connectivity and it does not change over time. All other within-community connectivities tend to increase with age. This agrees with the literature that large-scale brain functional modules tend to become more segregated with age, and as part of this process of segregation, the within-module connectivity strengthens \citep{fair2009functional}. There is also an age period, approximately around age 9 and 10, i.e., late childhood and early adolescence, where most within-community connectivity exhibit notable changes. On the other hand, most within-community connectivity remain unchanged starting from approximately age 13, i.e., adolescence. 

\begin{figure}[t!]
\centering
\includegraphics[scale=0.55,trim=10mm 20mm 0 10mm]{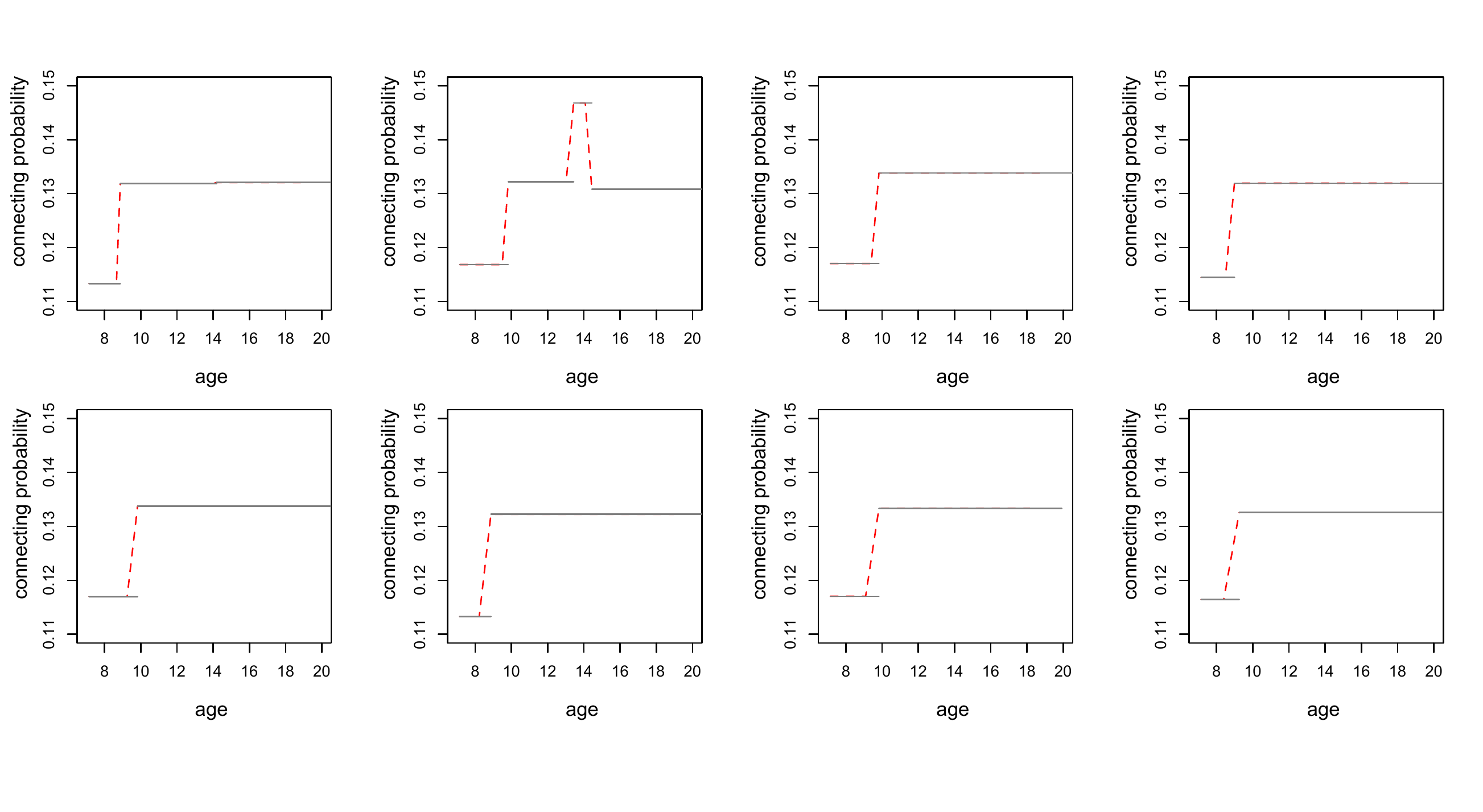}
\caption{The estimated connecting probability within the 4th community. The number of subjects in each interval from top to bottom and left to right is $n_S= \{10, 15, 20, 25, 30, 35, 40, 45\}$, respectively.}
\label{sen1}
\end{figure}

\begin{figure}[t!]
\centering
\includegraphics[scale=0.55,trim=10mm 20mm 0 10mm]{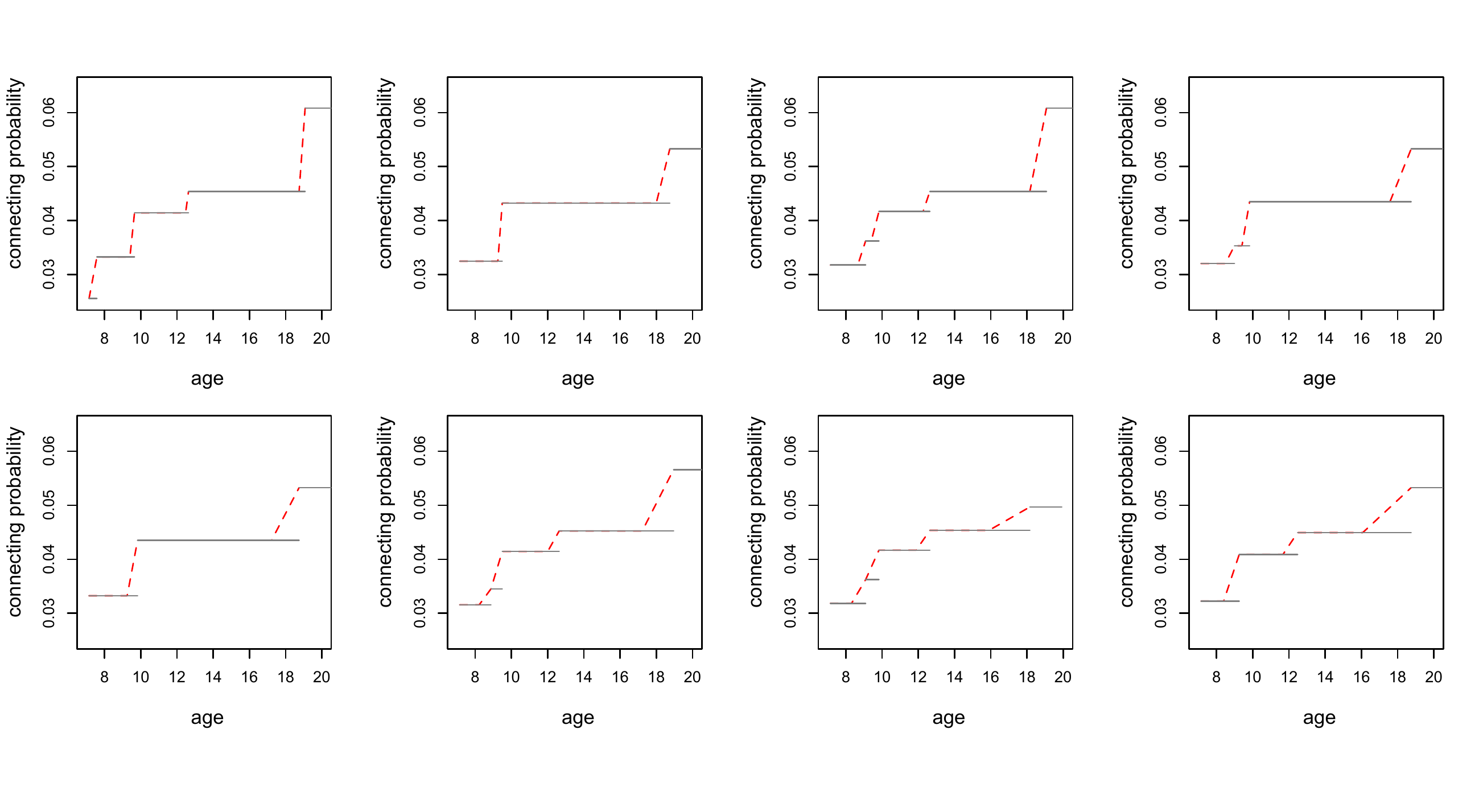}
\caption{The estimated connecting probability between the 9th and 10th communities. The number of subjects in each interval from top to bottom and left to right is $n_S= \{10, 15, 20, 25, 30, 35, 40, 45\}$, respectively.}
\label{sen2}
\end{figure}

Finally, we carried out a sensitivity analysis by varying the number of intervals $S$ so that the number of subjects in each interval ranges in $\{10, 15, 20, 25, 30, 35, 40, 45\}$, respectively. Our general finding is that the final estimates are not overly sensitive to the choice of $S$ as long as each interval contains a reasonably large number of subjects. As an illustration, we report in Figure \ref{sen1} the estimated within-community connectivity for the 4th module, \emph{default model network}, and report in Figure \ref{sen2} the between-community connectivity for the 9th and 10th modules, \emph{frontoparietal right} and \emph{frontoparietal left},  under different choices of $S$.

\end{document}